\documentclass[preprint, reprint, pra, showpacs,superscriptaddress]{revtex4-1}
\usepackage[latin1]{inputenc}
\usepackage{bm}
\usepackage{amssymb,amsbsy,amsmath}
\usepackage{graphicx}
\usepackage{verbatim}
\makeatletter
\usepackage{pifont}
\makeatother
\usepackage{dcolumn}
\usepackage{epstopdf}
\usepackage{subfigure}

\begin{document}

\title{Experimental generation of a high-fidelity four-photon linear cluster state}

\author{Chao Zhang}
\affiliation{Key Laboratory of Quantum Information, University of Science and Technology of China, CAS, Hefei, 230026, People's Republic of China}
\affiliation{Synergetic Innovation Center of Quantum Information and Quantum Physics, University of Science and Technology of China, Hefei, 230026, People's Republic of China}

\author{Yun-Feng Huang}
\email{hyf@ustc.edu.cn}
\affiliation{Key Laboratory of Quantum Information, University of Science and Technology of China, CAS, Hefei, 230026, People's Republic of China}
\affiliation{Synergetic Innovation Center of Quantum Information and Quantum Physics, University of Science and Technology of China, Hefei, 230026, People's Republic of China}

\author{Bi-Heng Liu}
\affiliation{Key Laboratory of Quantum Information, University of Science and Technology of China, CAS, Hefei, 230026, People's Republic of China}
\affiliation{Synergetic Innovation Center of Quantum Information and Quantum Physics, University of Science and Technology of China, Hefei, 230026, People's Republic of China}

\author{Chuan-Feng Li}
\email{cfli@ustc.edu.cn}
\affiliation{Key Laboratory of Quantum Information, University of Science and Technology of China, CAS, Hefei, 230026, People's Republic of China}
\affiliation{Synergetic Innovation Center of Quantum Information and Quantum Physics, University of Science and Technology of China, Hefei, 230026, People's Republic of China}

\author{Guang-Can Guo}
\affiliation{Key Laboratory of Quantum Information, University of Science and Technology of China, CAS, Hefei, 230026, People's Republic of China}
\affiliation{Synergetic Innovation Center of Quantum Information and Quantum Physics, University of Science and Technology of China, Hefei, 230026, People's Republic of China}

\date{\today}

\begin{abstract}
Cluster state plays a crucial role in the one-way quantum computation. Here, we propose and experimentally demonstrate a new scheme to prepare an ultrahigh-fidelity four-photon linear cluster state via spontaneous parametric down-conversion process. The state fidelity is measured to be $0.9517\pm0.0027$. Our scheme can be directly extended to more photons to generate N-qubit linear cluster state. Furthermore, our scheme is optimal for generating photonic linear cluster states in the sense of achieving the maximal success probability and having the simplest strategy. The key idea is that the photon pairs are prepared in some special non-maximally entangled states instead of the normal Bell states. To generate a 2N-qubit linear cluster state from N pairs of entangled photons, only (N-1) Hong-Ou-Mandel interferences are needed and a success probability of $(\frac{1}{4})^{N-1}$ is achieved.
\end{abstract}


\maketitle

{\section{Introduction}

Cluster states have attracted much interest because they are basic resource for the one-way quantum computation \cite{PhysRevLett.86.5188,PhysRevLett.86.910}. A one-way quantum computer works entirely different from the standard circuit model, because the computation in it proceeds by a sequence of single-qubit measurements on the pre-prepared cluster state with classical feedforward from the preceding measurement outcomes. Since single-qubit measurements are relatively easy to perform, the computational complexity mainly lies in the generation of the cluster state. Therefore, the ability to generate and manipulate such states becomes the most important requirement.

For photonic system, due to the lack of interactions between photons, a significant advantage of the one-way model is that the quantum gates in it can be implemented deterministically with classical feedforward via the cluster states, which is rather hard in the circuit model. In addition, the concept of the offline entangled state preparation gives us more space to simplify and optimize the generation schemes. For example, some methods based on cluster state quantum computation  \cite{PhysRevLett.95.010501,PhysRevLett.93.040503} have shown less resource requirement for many orders of magnitude than the original KLM scheme \cite{KnillLaflammeMilburn2001}.

Several experiments have been realized to prepare the four-photon linear cluster state  \cite{WaltherReschRudolphEtAl2005,KieselSchmidWeberEtAl2005,PhysRevLett.100.210501} through the spontaneous parametric down-conversion (SPDC) process. Some of them also give a proof-of-principle demonstration of the one-way quantum computer \cite{WaltherReschRudolphEtAl2005,PhysRevLett.100.210501}. Utilizing the multi-degree of freedom of photons, the qubit number has been enlarged to six, for example, a two-photon six-qubit \cite{PhysRevLett.103.160401} and a four-photon six-qubit linear cluster state \cite{PhysRevLett.104.020501} have been generated. Besides the linear cluster states, a kind of H-type six-photon cluster state \cite{LuZhouGuhneEtAl2007} and an eight-photon cluster state used for topological error correction \cite{YaoWangChenEtAl2012} have also been experimentally generated.

In this paper, we propose and experimentally realize a new scheme to prepare a four-photon linear cluster state. The key idea is to use non-maximally entangled state to remove the attenuation in the original scheme \cite{KieselSchmidWeberEtAl2005}. This modification dramatically increase the success probability from 1/9 to 1/4. To generate the non-maximally entangled state, we modify our ``sandwich-like" EPR source \cite{PhysRevLett.115.260402}, which has excellent qualities for multiphoton experiments. Using such sources, we obtain a four-photon linear cluster state with a fidelity up to $95\%$, which is much higher than any other previously reported results \cite{WaltherReschRudolphEtAl2005,KieselSchmidWeberEtAl2005,PhysRevLett.100.210501,PhysRevLett.98.180502,PhysRevLett.99.120503}. 

We also make a direct comparison with the original scheme, and experimentally demonstrate the superiority of our method. In addition, our scheme can be directly extended to larger linear cluster states. For example, to generate a 2N-qubit linear cluster state from N pairs of entangled photons, the new scheme can achieve a success probability of $(\frac{1}{4})^{N-1}$ with only (N-1) Hong-Ou-Mandel (HOM) interferences. Such a success probability has been proved to be optimal in a recent paper \cite{PhysRevA.91.062318} for generating the photonic linear cluster states from pairs of entangled photons. And the (N-1) HOM interferences obviously are the least number of interferences needed to connect N photon pairs, thus the new scheme implies a simplest strategy. Furthermore, although our experiment employs the SPDC system, our new scheme is not limited to the SPDC systems. In view of these, our new scheme would greatly benefit the generation of photonic linear cluster states.

\section{Experimental scheme}

Fig. 1(a) shows the original scheme of generating the four-photon linear cluster state. It starts from two Bell states $|\phi^+\rangle=\frac{1}{\sqrt{2}}(|HH\rangle+|VV\rangle)$, and performs a C-phase gate between the neighboring photons 2 and 3. When there is one and only one photon in each of the four output modes 1, 2', 3' and 4, the four photons are post-selected in the linear cluster state:
\begin{eqnarray}
\label{eq:targetstate}
|C_4\rangle &=& \frac{1}{2}(|HHHH\rangle_{12'3'4}+|HHVV\rangle_{12'3'4}\nonumber\\
& &+|VVHH\rangle_{12'3'4}-|VVVV\rangle_{12'3'4}),
\end{eqnarray}
where $|H\rangle$ and $|V\rangle$ represent the horizontal and vertical polarization states of the photons. The C-phase gate in the scheme is accomplished by three polarization dependent beam splitters (PDBSs). The essential interaction is realized by the central one. Its transmission efficiencies for $|H\rangle$ and $|V\rangle$ photons are set to be $T_H=1$ and $T_V=1/3$. So when the input state is $|VV\rangle$ (representing the logic state $|11\rangle$), the PDBS acts as a partial beam splitter and introduces a $\pi$ phase shift due to the second-order interference. When the input state is $|HH\rangle$, $|HV\rangle$ or $|VH\rangle$ (representing the logic states $|00\rangle$, $|01\rangle$ or $|10\rangle$), there is no interference on the PDBS and it only attenuates the V polarized photons. This effect is just like a C-phase gate. In order to equalize the parameters for each output components, two additional PDBSs with complementary transmissions ($T_H=1/3$, $T_V=1$) are placed at the two output ports. The success probability to find a coincidence at the final four output modes is 1/9.

As the additional PDBSs only act as local polarization compensators. They can be equally placed in the input side (as shown in Fig. 1(b)). Then, we can further integrate them into the source. So, our scheme (Fig. 1(c)) employs the non-maximally entangled states as input to remove the attenuation of the H polarizations, which will significantly increase the success probability. The input state is
\begin{eqnarray}
\label{eq:input}
|\psi^{(in)}\rangle &=&\biggl(\frac{1}{2}|HH\rangle+\frac{\sqrt{3}}{2}|VV\rangle\biggr)_{12}\biggl(\frac{1}{2}|HH\rangle+\frac{\sqrt{3}}{2}|VV\rangle\biggr)_{34}\nonumber\\
&=& \frac{1}{4}(h^\dag_1h^\dag_2h^\dag_3h^\dag_4+\sqrt{3}h^\dag_1h^\dag_2v^\dag_3v^\dag_4+\sqrt{3}v^\dag_1v^\dag_2h^\dag_3h^\dag_4+\nonumber\\
& &3v^\dag_1v^\dag_2v^\dag_3v^\dag_4)|0\rangle,
\end{eqnarray}
The mode transformation of the central PDBS is
\begin{equation}
\left\{
\begin{aligned}
\label{eq:PDBS}
h^\dag_2 &\rightarrow h^\dag_{2'}\\
h^\dag_3 &\rightarrow h^\dag_{3'}\\
v^\dag_2 &\rightarrow \sqrt{\frac{1}{3}} v^\dag_{2'} + i\sqrt{\frac{2}{3}}v^\dag_{3'}\\
v^\dag_3 &\rightarrow \sqrt{\frac{1}{3}} v^\dag_{3'} + i\sqrt{\frac{2}{3}}v^\dag_{2'}
\end{aligned}
\right.
\end{equation}
where $h^\dag_{i(i')}$ and $v^\dag_{i(i')}$ denote the creation operators for the horizontal and vertical polarized photons in the $i(i')-th$ spatial mode and the $|0\rangle$ denotes the vacuum state.
The output state can be separated into two parts \cite{PhysRevA.91.062318}:
\begin{equation}
\label{eq:outputth}
|\psi^{(out)}\rangle=\alpha|\psi^{(out)}_{\uppercase\expandafter{\romannumeral1}}\rangle+\beta|\psi^{(out)}_{\uppercase\expandafter{\romannumeral2}}\rangle, |\alpha|^2+|\beta|^2=1.
\end{equation}
where $|\psi^{(out)}_{\uppercase\expandafter{\romannumeral1}}\rangle$ represents the cases that there is one and only one photon in each output port, while $|\psi^{(out)}_{\uppercase\expandafter{\romannumeral2}}\rangle$ represents the other cases. So the output state is
\begin{eqnarray}
\label{eq:output}
|\psi^{(out)}\rangle &=& \frac{1}{4}(|HHHH\rangle+|HHVV\rangle+|VVHH\rangle \nonumber\\
                     & & -|VVVV\rangle)_{12'3'4}+\frac{\sqrt{3}}{2}|\psi^{(out)}_{\uppercase\expandafter{\romannumeral2}}\rangle \nonumber\\
                     &=&\frac{1}{2}|C_4\rangle+\frac{\sqrt{3}}{2}|\psi^{(out)}_{\uppercase\expandafter{\romannumeral2}}\rangle.
\end{eqnarray}
Thus the success probability is equal to $|\alpha|^2=1/4$.

We also notice another scheme to generate the four-photon cluster state $|C_4\rangle$ \cite{PhysRevLett.100.210501}, which uses only HWPs and PBSs. The success probability is also 1/4. The advantage of this scheme is that it starts with only one entangled photon pair and the other two photons are initially prepared in a product state. This will make the experiment setup of the input part simpler. However, such a simplification also results in the requirement of two HOM interferences in their scheme, which makes it more complex in the latter part. As a result, it can only keep one photon as trigger. Here, the word trigger means that by detecting this photon we can be sure about the generation of at least one photon pair from the corresponding SPDC source, because the trigger photon's spatial mode does not overlap with any other photon's spatial mode. For SPDC systems, which are still the preferred system for multi-photon experiments, less trigger photons will lead to more higher-order photon pair emission noise  \cite{YaoWangXuEtAl2012a,HuangLiuPengEtAl2011}. More importantly, such scheme cannot be directly extended for generating a six-photon cluster state with the SPDC process, since it will introduce the same-order noise, unless a PDBS is used for connecting the third photon pair. Here, the same-order noise means that by post-selecting six-fold coincidences in such optical scheme one can not remove all the unwanted cases of the same three-photon-pair-order emission, for example, the case of the first SPDC source emitting two photon pairs, the second emitting one pair, while the third emitting no pair would still result in six-fold coincidence counts, which will ruin the experiment results.

\begin{figure}[tb]
\centering
\includegraphics[width=0.45\textwidth]{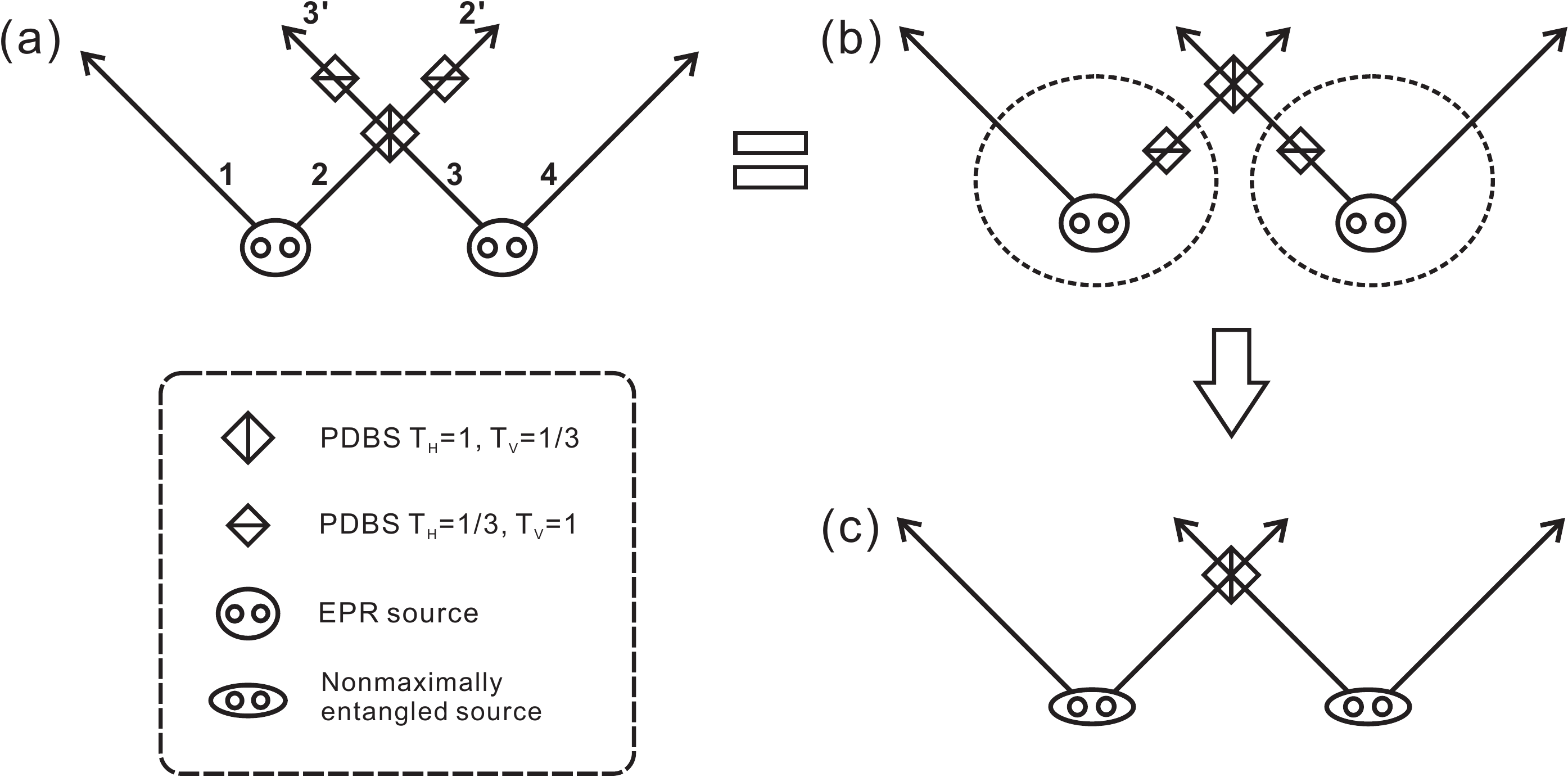}
\caption{\label{Fig:1} (a) The original scheme to prepare the four-photon linear cluster state. (b) It is equal to put the additional PDBSs in the input side. (c) Our new scheme uses non-maximally entangled states as input. }  \end{figure}

\section{Experimental setup and results}

\begin{figure}[tb]
\centering
\includegraphics[width=0.45\textwidth]{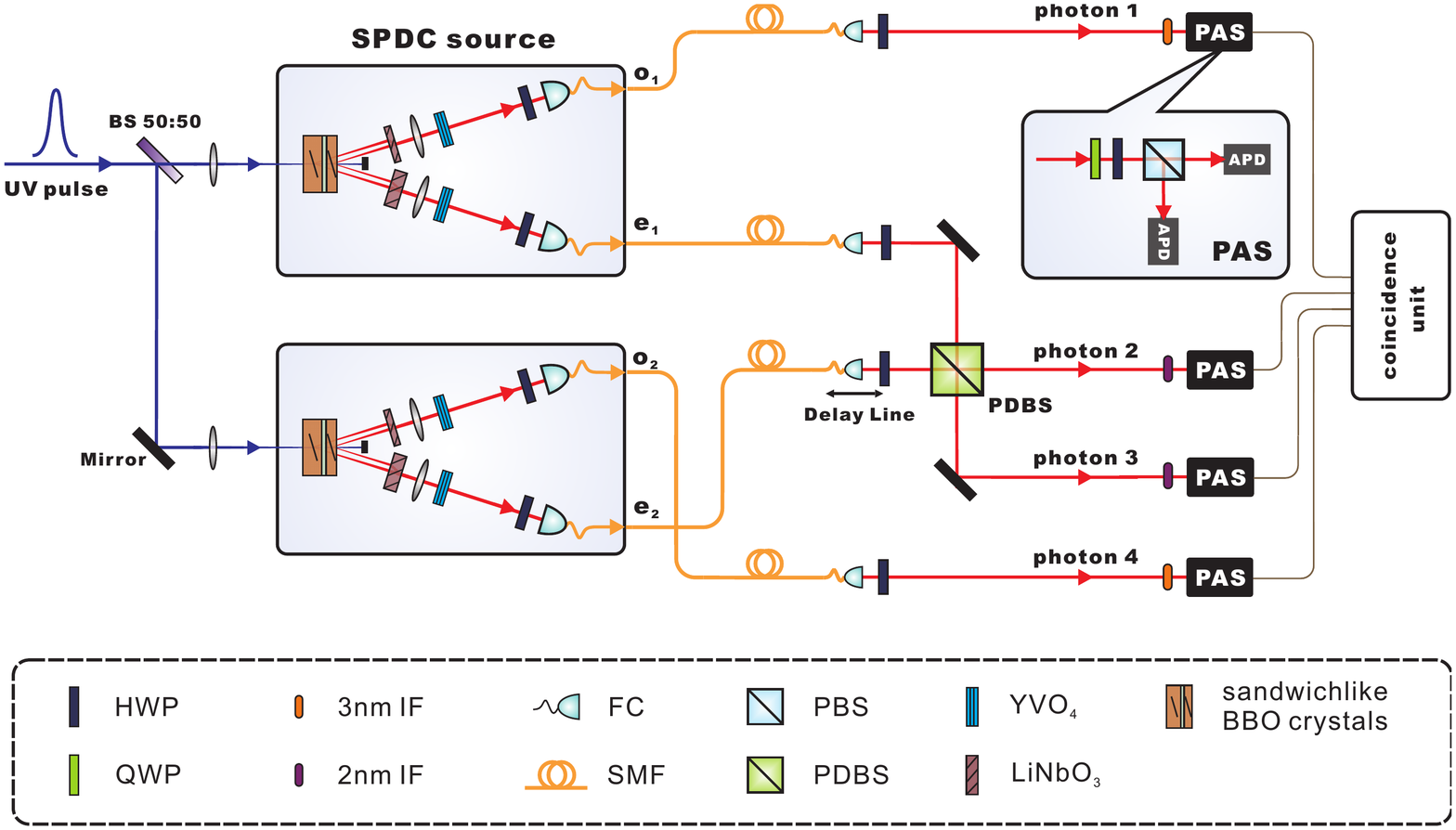}
\caption{\label{Fig:2} The experimental setup. An ultrafast UV pump pulse frequency doubled from a Ti:Sapphire mode-locked laser (with a central wavelength of 780 nm, a pulse duration of 140 fs and a repetition rate of 76 MHz) is evenly separated into two beams, and subsequently sent to pump two sandwich-like SPDC sources (as shown in the two left boxes). The two sources both produce the state $\frac{1}{2}|HH\rangle+\frac{\sqrt{3}}{2}|VV\rangle$. Each source uses two BBO crystals with 2 mm and 1 mm thicknesses. The LiNbO$_3$ crystals are used for spatial compensation with 0.5 mm and 4.2 mm thicknesses for o- and e-photons. The YVO$_4$ crystals are used for temporal compensation with 0.57 mm and 0.47 mm thicknesses for o- and e-photons. The YVO$_4$ crystals can also be used for tuning the relative phase between the four terms in the cluster state. Each photon is finally measured by a polarization analyzing system (PAS), the details are shown in the right inset. }  \end{figure}

The experimental setup is shown in Fig. 2. The two SPDC sources are specially designed to prepare the nonmaximally entangled states $|\psi_s\rangle=\frac{1}{2}|HH\rangle+\frac{\sqrt{3}}{2}|VV\rangle$. The down-converted photons are coupled into the single-mode fibers (SMFs) for spatial filtering. The output of the two e(extraordinary)-photons are directed to interfere on the PDBS. The PDBS is set as $T_H=1, T_V=1/3$. The transmission rate of the V-polarized photons can be finely adjusted by tuning the photon incident angle on the PDBS. One of the output couplers is mounted on a translation stage as a delay line to adjust the photon arriving time. The zero delay is determined by the HOM interference dip of two V-polarized photons, which means that they arrive at the PDBS simultaneously. To make the four terms in the target state indistinguishable, the time difference information between the H-polarized and V-polarized photons in each output port (before polarization analyzing systems) should be erased. This is done by the temporal compensation crystals in the two SPDC sources. However, the pump pulse duration introduces a time jitter of the down-converted photons between independent SPDC sources. This time information can not be erased simply by the path delay or compensation crystals. We can use narrow-band spectral filters to increase the coherence length of the interfering photons, and thus increase their indistinguishability. In the experiment, we use 2 nm (FWHM) bandwidth interference filters (IF) for interfering (extraordinary) photons and 3 nm (FWHM) IFs for trigger (ordinary) photons. After all the cases are made indistinguishable, the target state can be successfully projected by post-selection.

The entanglement source \cite{PhysRevLett.115.260402} we used to generate $|\psi_s\rangle$ has a sandwich-like structure: one true-zero-order half-wave plate (THWP) sandwiched by two type-II beamlike phase-matching \cite{KurtsieferOberparleiterWeinfurter2001} $\beta$-barium borate (BBO) crystals. However, the original source uses two 1-mm-thick BBO crystals and can only produce the Bell state. In our work, we find that the photon pair generation rate of a 2-mm-thick BBO crystal is nearly three times as that of a 1-mm-thick crystal, but not two times. The reason is that the thicker the crystal, the narrower the spectral width of the down-converted photons will be. So the counting rate does not linearly increase with the crystal thickness under fixed narrow-band spectral filtering are used. This $\sim 3:1$ ratio of the photon pair generation rates between the two BBO crystals can be further finely adjusted by tuning the fiber coupling. Thus, we used a 2-mm-thick and a 1-mm-thick BBO in the sandwich structure to generate $|\psi_s\rangle$. When the two possible ways of generating photon pairs (through the first or the second BBO crystal) are made indistinguishable by spatial and temporal compensations, the state $|\psi_s\rangle$ is prepared. The brightness of our source is 12000 pairs/s with 40 mW pump power (using 2 nm and 3 nm bandwidth IFs), which is much higher than the type-I entanglement source \cite{PhysRevA.60.R773}, due to its ``beamlike" emission characteristics.

After preparation, each photon is measured by a polarization analyzing system (PAS), which consists of one quarter-wave plate (QWP), one half-wave plate (HWP), one polarizing beam splitter (PBS), and two avalanche photodiodes (APDs). A programmable multi-channel coincidence unit is used to register all possible coincidence events, and we only post-select the four-fold coincidence when there is one and only one APD firing in each of the four PAS. In the experiment, we use 40 mW pump power for each SPDC source and the four-photon cluster state generation rate is 0.5 Hz.

\begin{table}[tb]
\centering
\caption{\label{Table:1} Stabilizer correlations. The 16 expectation values can be obtained from 9 joint measurement settings (see Appendix A for details) consisting of local measurements of each photon. The error bars are deduced from the raw data (see Appendix A) and poisson counting statistics.}
\begin{tabular}{c c c c} \hline

 &Stabilizer&Operators&Expectation Value\\\hline
(1)    &$g_1$             &$Z_1Z_2I_3I_4$     &$0.9864\pm0.0032$ \\
(2)    &$g_2$             &$X_1X_2Z_3I_4$     &$0.9474\pm0.0113$ \\
(3)    &$g_3$             &$I_1Z_2X_3X_4$     &$0.9290\pm0.0129$ \\
(4)    &$g_4$             &$I_1I_2Z_3Z_4$     &$0.9773\pm0.0041$ \\
(5)    &$g_1g_2$          &$-Y_1Y_2Z_3I_4$     &$0.9646\pm0.0091$ \\
(6)    &$g_1g_3$          &$Z_1I_2X_3X_4$     &$0.9315\pm0.0127$ \\
(7)    &$g_1g_4$          &$Z_1Z_2Z_3Z_4$     &$0.9773\pm0.0113$ \\
(8)    &$g_2g_3$          &$X_1Y_2Y_3X_4$     &$0.9342\pm0.0132$ \\
(9)    &$g_2g_4$          &$X_1X_2I_3Z_4$     &$0.9474\pm0.0137$ \\
(10)   &$g_3g_4$          &$-I_1Z_2Y_3Y_4$     &$0.9301\pm0.0091$ \\
(11)   &$g_1g_2g_3$       &$Y_1X_2Y_3X_4$     &$0.9261\pm0.0137$ \\
(12)   &$g_1g_2g_4$       &$-Y_1Y_2I_3Z_4$     &$0.9646\pm0.0091$ \\
(13)   &$g_1g_3g_4$       &$-Z_1I_2Y_3Y_4$     &$0.9249\pm0.0137$ \\
(14)   &$g_2g_3g_4$       &$X_1Y_2X_3Y_4$     &$0.9445\pm0.0122$ \\
(15)   &$g_1g_2g_3g_4$    &$Y_1X_2X_3Y_4$     &$0.9429\pm0.0123$ \\
(16)   &$I$               &$I_1I_2I_3I_4$     &1.0 \\\hline
 & & &$F_{C_4}=0.9517\pm0.0027$ \\\hline

\end{tabular}
\end{table}

In order to characterize the prepared four-photon cluster state, we obtain the fidelity of it by $F=Tr(|C_4\rangle \langle C_4|\rho_{exp})$, where $\rho_{exp}$ denotes the experimentally prepared state. For graph states, their projector operator (Here is the $|C_4\rangle \langle C_4|$) can be completely described by their stabilizers \cite{hein2006entanglement}. The target state $|C_4\rangle$ corresponds to the standard four-qubit linear cluster state after Hadamard transformations in the qubits 1 and 4. Thus, the stabilizing operators are:
\begin{eqnarray}
\label{eq:stabilizer}
g_1 &=& Z_1Z_2I_3I_4 \nonumber\\
g_2 &=& X_1X_2Z_3I_4 \nonumber\\
g_3 &=& I_1Z_2X_3X_4 \nonumber\\
g_4 &=& I_1I_2Z_3Z_4
\end{eqnarray}
where the subscript i labels the qubits and $X_i$, $Y_i$, and $Z_i$ denote the Pauli operators $\sigma_x$, $\sigma_y$ and $\sigma_z$. The target state $|C_4\rangle$ is uniquely defined by $g_i|C_4\rangle=|C_4\rangle (i=1...4)$. These operators $g_i$ also form a group called $\emph{stabilizer}$ (the S group), which consists of themselves and their products. The projector can be written as the average of the $2^N$ elements in the S group: $|C_4\rangle \langle C_4|=\frac{1}{2^4}\sum_{\sigma\in S}\sigma$. Therefore, the fidelity for the target state equals to the average expectation value of all the $\emph{stabilizer}$ operators. The measurement result is shown in Table I. The state fidelity is calculated to be $F_{C_4}=0.9517\pm0.0027$. Our result clearly demonstrates the genuine four-photon entanglement with 167 standard deviations according to the projector-based witness. And it is also high enough to distinguish the observed state from other types of genuine four-qubit entanglement, e.g., GHZ type and W type entanglement \cite{PhysRevA.74.020301}. The 15 nontrivial stabilizer correlations can also be used for nonlocality test, since they can be used to construct GHZ-type arguments \cite{PhysRevA.71.042325}. For example, the four correlations:
\begin{eqnarray}
\label{eq:argument}
(2)&\  X_1X_2Z_3I_4 \ &= 1\nonumber\\
(5)&\  Y_1Y_2Z_3I_4 \ &= -1\nonumber\\
(8)&\  X_1Y_2Y_3X_4 \ &= 1\nonumber\\
(11)&\ Y_1X_2Y_3X_4 \ &= 1
\end{eqnarray}
can deduce a contradiction between the local realism and the quantum theory. In our measurements, the largest error rate of data flipping is $0.0376\pm0.0068$ (corresponding to the $Z_1I_2Y_3Y_4$ measurement), which is well below the threshold of $\frac{1}{4}$ \cite{:/content/aapt/journal/ajp/65/12/10.1119/1.18758}. Thus our results demonstrate the nonlocality of the state clearly.

\section{Conclusion}

\begin{figure}[tb]
\centering
\includegraphics[width=0.45\textwidth]{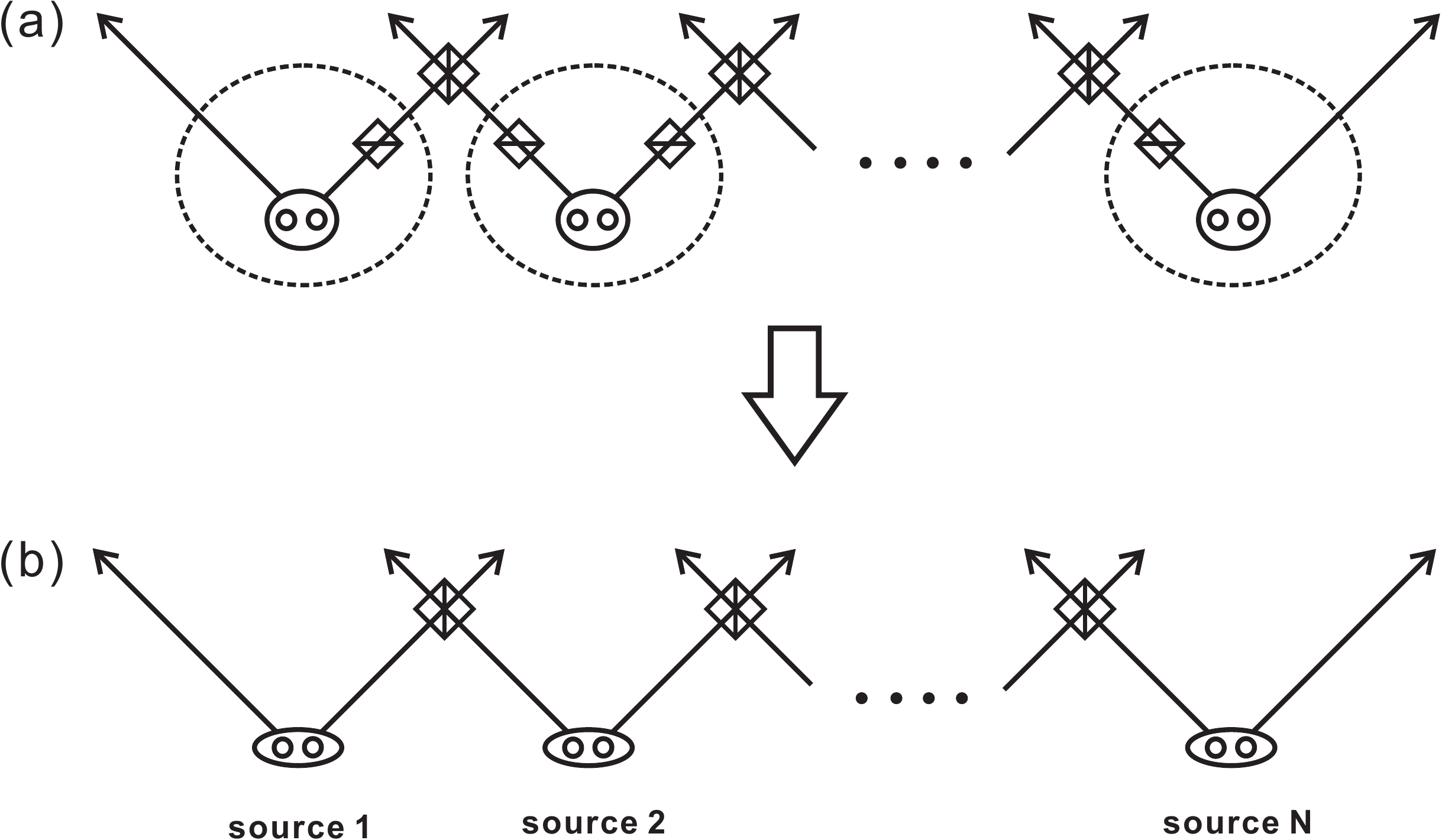}
\caption{\label{Fig:3} (a) Original scheme to prepare 2N-qubit linear cluster state. (b) Our new scheme with non-maximally entangled states as input. }  \end{figure}

Our scheme can be easily generalized to larger linear cluster state, as shown in Fig. 3. The original scheme uses C-phase gates (three PDBSs) to connect N photon pairs in Bell states, which directly reflects the definition of the graph states. Now we can integrate the attenuation PDBSs into the sources. As shown in Fig. 3(a), the parts in the dotted line circle can be replaced by some special non-maximally entangled states. The source in the beginning or the end has one attenuation PDBS, so the required non-maximally entangled state is $|\psi_s\rangle$. The other sources all have two attenuation PDBSs, so the required state is $\frac{1}{4}|HH\rangle+\frac{\sqrt{3}}{4}|HV\rangle+\frac{\sqrt{3}}{4}|VH\rangle-\frac{3}{4}|VV\rangle$. Using the Schmidt decomposition \cite{nielsen2010quantum}, the state is equal to $\frac{\sqrt{7}-1}{4}|HH\rangle-\frac{\sqrt{7}+1}{4}|VV\rangle$ through single qubit rotation. So we need only two kinds of non-maximally entangled sources. To connect them into a 2N-qubit linear cluster state, as shown in Fig. 3(b), we need only (N-1) HOM interferences. And the success probability is $(\frac{1}{4})^{N-1}$ (see Appendix B for details), which is equal to the maximal success probability of transforming N photon pairs in Bell states to a 2N-qubit linear cluster state \cite{PhysRevA.91.062318}. However, the methods introduced in \cite{PhysRevLett.100.210501,PhysRevA.91.062318} need more HOM interferences and may introduce the same-order noise if the SPDC process is employed. Furthermore, it is obvious that connecting N photon pairs into one linear cluster chain requires at least (N-1) HOM interferences, so our method provides a simplest strategy to achieve the optimal success probability.

In conclusion, we propose a new method to prepare photonic linear cluster states using non-maximally entangled states as input. And we also experimentally generate a rather high quality four-photon linear cluster state. We also show that our scheme is more efficient than previous methods in both the success probability and the required number of second-order interferences. Combined with quantum memory and the recycling technique, our method would be useful as a building block for large scale photonic cluster state generation.

\begin{acknowledgments}
This work was supported by the National Natural Science Foundation of China (No. 61327901, No. 61490711, No. 11274289, No. 11325419, No. 61225025, No. 11474268, No. 11374288), the Strategic Priority Research Program (B) of the Chinese Academy of Sciences (Grant No. XDB01030300), the National Program for Support of Top-notch Young Professionals (No. BB2470000005), and the Fundamental Research Funds for the Central Universities (No. WK2470000018, No. WK2470000022).
\end{acknowledgments}

\begin{widetext}
\section *{Appendix A: Measurement settings and raw data for the 16 stabilizer correlations}

The nine joint measurement settings employed in our experiment are shown as below (the subscripts of particles are omitted) \cite{PhysRevA.74.020301}:
\begin{equation*}
ZZZZ, ZZXX, XXZZ, ZZYY, YYZZ, XYXY, XYYX, YXXY, YXYX.
\end{equation*}
We can use $ZZZZ$ to obtain the expectation values of the operators $\{ZZII, IIZZ, ZZZZ\}$, $ZZXX$ for $\{IZXX, ZIXX\}$, $XXZZ$ for $\{XXZI, XXIZ\}$, $ZZYY$ for $\{IZYY, ZIYY\}$, and $YYZZ$ for $\{YYZI, YYIZ\}$. The raw data for the nine local measurement settings is shown in Table A1.

\begin{table}[htb]
\centering
\setcounter{table}{0}    
\renewcommand{\thetable}{A\arabic{table}}
\caption{\label{Table:1} The raw data for the nine local measurement settings. All the 16 possible combinations of the fourfold coincidence events are recorded. Here we use ``0 (1)" denote the H (V) port detector firing in each output mode. The data collection time for the ZZZZ measurement is 5000s, for other measurement settings is 1500s. In the last column we calculate the error rate (defined as the proportion of the incorrect terms) for each measurement setting.}
\begin{tabular}{|c| c c c c c c c c c c c c c c c c |c|} \hline

Setting&0000&0001&0010&0011&0100&0101&0110&0111&1000&1001&1010&1011&1100&1101&1110&1111&Error Rate\\\hline
ZZZZ&577&2&6&639&2&5&0&3&0&0&4&4&664&6&7&721&$0.0148\pm0.0023$\\
ZZXX&208&1&0&171&4&2&1&4&3&0&1&2&10&195&206&9&$0.0453\pm0.0073$\\
XXZZ&189&1&3&5&0&0&0&173&0&0&1&219&189&5&4&9&$0.0355\pm0.0066$\\
ZZYY&1&151&190&2&0&0&0&1&0&0&1&0&213&13&11&188&$0.0376\pm0.0068$\\
YYZZ&0&0&3&232&206&0&1&4&208&1&1&5&2&2&0&183&$0.0229\pm0.0052$\\
XYXY&90&2&5&83&1&83&76&3&2&78&91&2&103&4&1&97&$0.0277\pm0.0061$\\
XYYX&108&2&3&67&3&93&96&4&2&98&94&2&102&8&2&106&$0.0329\pm0.0063$\\
YXXY&93&1&1&76&2&80&86&5&6&93&107&2&94&3&1&85&$0.0286\pm0.0061$\\
YXYX&94&2&5&76&5&89&90&1&5&97&87&3&87&2&5&110&$0.0369\pm0.0069$\\\hline
\end{tabular}
\end{table}



\section *{Appendix B: Calculation of the success probability}

First we demonstrate the standard n-qubit linear cluster state has the following recurrence relation \cite{PhysRevA.91.062318}:
\begin{equation}
\setcounter{equation}{1}    
\renewcommand{\theequation}{A\arabic{equation}}
|C_n\rangle=\frac{1}{\sqrt{2}}(|C_{n-1}\rangle|H\rangle_n+|\tilde{C}_{n-1}\rangle|V\rangle_n).
\end{equation}
Here, $|C_n\rangle$ represents the n-qubit linear cluster state and $|\tilde{C}_{n-1}\rangle=\sigma_z^{n-1}|C_{n-1}\rangle$, while $|H\rangle$ and $|V\rangle$ denote the logic $|0\rangle$ and $|1\rangle$ states, $\sigma_z^{n-1}$ is the Pauli z matrix of the qubit (n-1).

Obviously,
\begin{equation*}
|C_2\rangle=\frac{1}{\sqrt{2}}(|+\rangle|H\rangle+|-\rangle|V\rangle)=\frac{1}{\sqrt{2}}(|C_1\rangle|H\rangle+|\tilde{C}_1\rangle|V\rangle)
\end{equation*}
For $n=k\ (k>1)$, assume the Eq. A1 holds
\begin{equation*}
|C_k\rangle=\frac{1}{\sqrt{2}}(|C_{k-1}\rangle|H\rangle_k+|\tilde{C}_{k-1}\rangle|V\rangle_k)
\end{equation*}
Then according to the definition of the graph state, we have
\begin{eqnarray*}
|C_{k+1}\rangle &=& \mathrm{CPhase}_{(k,k+1)}|C_k\rangle\otimes|+\rangle_{k+1}\\
                &=& \mathrm{CPhase}_{(k,k+1)}\frac{1}{\sqrt{2}}(|C_{k-1}\rangle|H\rangle_k+|\tilde{C}_{k-1}\rangle|V\rangle_k)\otimes\frac{1}{\sqrt{2}}(|H\rangle+|V\rangle)_{k+1}\\
                &=& \frac{1}{2}(|C_{k-1}\rangle|HH\rangle+|C_{k-1}\rangle|HV\rangle+|\tilde{C}_{k-1}\rangle|VH\rangle-|\tilde{C}_{k-1}\rangle|VV\rangle)\\
                &=& \frac{1}{\sqrt{2}}(|C_k\rangle|H\rangle_{k+1}+|\tilde{C}_k\rangle|V\rangle_{k+1})
\end{eqnarray*}
So we demonstrate the recurrence relation of Eq. A1. Note that we use the standard definition of the linear cluster state here, according to which the four-qubit linear cluster state $|C_4\rangle$ is locally equivalent, under Hardmard transformations performed on qubit 1 and 4, to the one defined in the main text, i. e., the state $\frac{1}{2}(|HHHH\rangle+|HHVV\rangle+|VVHH\rangle-|VVVV\rangle)$.

Next we calculate the output state for Fig. 4(b) in the main text. The produced states for source 1 and source N are $\frac{1}{2}|+H\rangle+\frac{\sqrt{3}}{2}|-V\rangle$ and $\frac{1}{2}|H+\rangle+\frac{\sqrt{3}}{2}|V-\rangle$ respectively, for the sources in the middle are
$\frac{1}{4}|HH\rangle+\frac{\sqrt{3}}{4}|HV\rangle+\frac{\sqrt{3}}{4}|VH\rangle-\frac{3}{4}|VV\rangle$. Using such sources, we can demonstrate the output state for the preceding $k\ (1\leq k\leq N-1)$ pairs of entangled photons after connections is
\begin{equation}
\renewcommand{\theequation}{A\arabic{equation}}
\label{eq:koutput}
|\psi^{out}_{2k}\rangle=\frac{1}{2}|C_{2k-1}\rangle|H\rangle_{2k}+\frac{\sqrt{3}}{2}|\tilde{C}_{2k-1}\rangle|V\rangle_{2k}.
\end{equation}
For $k=1$, we have
\begin{equation*}
|\psi^{out}_{2}\rangle=\frac{1}{2}|+\rangle|H\rangle+\frac{\sqrt{3}}{2}|-\rangle|V\rangle=\frac{1}{2}|C_{1}\rangle|H\rangle+\frac{\sqrt{3}}{2}|\tilde{C}_{1}\rangle|V\rangle
\end{equation*}
For $k=2$, we have
\begin{eqnarray*}
|\psi^{out}_4\rangle &=& \mathbf{U}_{\mathrm{PDBS}}^{2,3}\biggl(\frac{1}{2}|C_1\rangle|H\rangle+\frac{\sqrt{3}}{2}|\tilde{C}_1\rangle|V\rangle\biggr)_{1,2}\otimes\biggl(\frac{1}{4}|HH\rangle+\frac{\sqrt{3}}{4}|HV\rangle+\frac{\sqrt{3}}{4}|VH\rangle-\frac{3}{4}|VV\rangle\biggr)_{3,4}\\
                    &=& \frac{1}{8}\bigl(|C_1HHH\rangle+\sqrt{3}|C_1HHV\rangle+|C_1HVH\rangle-\sqrt{3}|C_1HVV\rangle+|\tilde{C}_1VHH\rangle+\sqrt{3}|\tilde{C}_1VHV\rangle \\
                    & & -|\tilde{C}_1VVH\rangle+\sqrt{3}|\tilde{C}_1VVV\rangle\bigr)+others \\
                    &=& \frac{\sqrt{2}}{8}\bigl(|C_1H+\rangle+|\tilde{C}_1V-\rangle\bigr)|H\rangle+\frac{\sqrt{6}}{8}\bigl(|C_1H-\rangle+|\tilde{C}_1V+\rangle\bigr)|V\rangle+others \\
                    &=& \frac{1}{2}\biggl(\frac{1}{2}|C_3\rangle|H\rangle+\frac{\sqrt{3}}{2}|\tilde{C}_3\rangle|V\rangle\biggr)+others
\end{eqnarray*}
Here we use $\mathbf{U}_{\mathrm{PDBS}}^{i,j}$ to represent the mode transformation matrix of the PDBS (Eq. (3) in the main text) for the modes i, j, and we use ``others" to represent the failed cases of the connection ($|\psi_{II}^{(out)}\rangle$ in the Eq. (4) of the main text). In the last step we use the recurrence relation of the cluster state. Continue to iterate this process we can calculate the output state for $k=N-1$:
\begin{equation*}
|\psi^{out}_{2N-2}\rangle=\biggl(\frac{1}{2}\biggr)^{N-2}\biggl(\frac{1}{2}|C_{2N-3}\rangle|H\rangle_{2N-2}+\frac{\sqrt{3}}{2}|\tilde{C}_{2N-3}\rangle|V\rangle_{2N-2}\biggr)+others
\end{equation*}
For the last connection
\begin{eqnarray*}
|\psi^{out}_{2N}\rangle &=& \biggl(\frac{1}{2}\biggr)^{N-2} \mathbf{U}_{\mathrm{PDBS}}^{2N-2,2N-1}\biggl(\frac{1}{2}|C_{2N-3}\rangle|H\rangle_{2N-2}+\frac{\sqrt{3}}{2}|\tilde{C}_{2N-3}\rangle|V\rangle_{2N-2}\biggr)\otimes\biggl(\frac{1}{2}|H\rangle|+\rangle+\frac{\sqrt{3}}{2}|V\rangle|-\rangle\biggr)_{2N-1,2N} \\
&=&\biggl(\frac{1}{2}\biggr)^{N-2}\frac{1}{4}\bigr(|C_{2N-3}\rangle|HH+\rangle+|C_{2N-3}\rangle|HV-\rangle+|\tilde{C}_{2N-3}\rangle|VH+\rangle-|\tilde{C}_{2N-3}\rangle|VV-\rangle\bigr)+others\\
&=&\biggl(\frac{1}{2}\biggr)^{N-1}|C_{2N}\rangle+others
\end{eqnarray*}
So we demonstrate that the output state is indeed the 2N-qubit linear cluster state after post-selection. The coefficient $(\frac{1}{2})^{N-1}$ before the state vector indicate the success probability for post-selection is $(\frac{1}{4})^{N-1}$, which means that the success probability for every connection is equal to $\frac{1}{4}$. The reason is that every PDBS loses three fourths of the photons according to its transformation matrix and the input states.

\end{widetext}


%


\end{document}